\documentclass{article}
\usepackage{amsfonts}

\usepackage{graphicx}
\usepackage{amsmath}

\input{tcilatex}

\begin{document}

\title{Thermostatistics of the G-on Gas}
\author{F.B\"{u}y\"{u}kk\i l\i \c{c}$\thanks{%
e-mail: fevzi@sci.ege.edu.tr}$, H.Uncu, D.Demirhan \\
Department of Physics,Science Faculty,Ege\\
University,35100 Bornova, Izmir, Turkey .}
\maketitle

\begin{abstract}
In this study, the particles of the quantum gases, namely bosons and
fermions are regarded as g-ons by the parameter of the fractional exclusion
statistics g, where $0\leq g\leq 1$. With this point of departure, the
distribution function of the g-on gas is obtained by the variational,
steepest descent and statistical methods. The distribution functions which
are found by means of these three methods are compared. It is seen that the
thermostatistical formulations of quantum gases can be unified. By suitable
choices of g standard relations of statistical mechanics of the Bose and
Fermi systems are recovered.
\end{abstract}

\bigskip \bigskip {\huge Introduction}

Particles of quantum systems, including well known fermions and bosons can
be called as g-ons which obey fractional exclusion statistics. The aim of
this study is to devolop the distribution function of a g-on gas by means of
variational, steepest descent and statistical methods and discuss them.
Futhermore, another purpose is \ to unify the different thermostatistical
formulations of quantum gases where the concomitant distribution function of
g-on gas is used.

The outline of this paper is as follows: In section 1.1, Haldane statistics
is obtained by the variational method where the entropy of the g-on gas is
employed. By means of an approximate solution to Haldane statistics, an
elegant formula for the distribution function of the g-on gas is found. In
section 1.2, where the steepest descent method is applied to the partition
function, an exact solution of distribution function is obtained. In section
1.3, the probability of occupation numbers and average distribution
functions are investigated by the statistical method. In section 2 and 3,
fluctuations in the occupation number and thermodynamics of the g-on gas are
respectively discussed.

\section{Derivation of the distribution function of the g-on gas system}

\subsection{Variational method}

\qquad Let us consider an ideal g-on gas system which consists of N
particles. Let\ the system be in heat and particle baths that evolves
towards equilibrium, that is, the total energy E and the total number of
g-ons N fluctuate, \ but, these quantities are conserved on the average. The
entropy of the ideal g-on system which is not in equilibrium has been
obtained from the statistical weight and is given by [1],

\begin{equation}
S_{g}=\sum_{k=1}^{K}g_{k}\left[ 1+n_{k}\left( 1-g\right) \right] \ln
[1+n_{k}\left( 1-g\right) -\left( 1-gn_{k}\right) \ln \left[ 1-gn_{k}\right]
-n_{k}\ln n_{k}
\end{equation}
where the Boltzmann constant is set to unity, $g_{k}$ is the degeneracy, $%
n_{k}$ is the occupation number, K is the total number of states; \ g is the
parameter that characterizes the complete or partial \ action of the
exclusion principle i.e. makes interpolation between quantum gases [2,3].
Following the variational method outlined in the  ref.[1], the following
equation is obtained 
\begin{equation}
\left[ \ln 1+n_{k}\left( 1-g\right) \right] ^{\left( 1-g\right) }+\left(
1-g\right) +\ln \left( 1-gn_{k}\right) ^{g}+g-\ln n_{k}-1-\alpha -\beta
\varepsilon _{k}=0
\end{equation}
where $\alpha =-\frac{\mu }{T}$\bigskip\ , $\beta =\frac{1}{T}$ , $%
\varepsilon _{k}$ is is the energy of the k. state and $\mu $ is the
chemical potential. After performing some algebra Eq.(2) becomes; 
\begin{equation}
\frac{\left[ 1+n_{k}\left( 1-g\right) \right] ^{1-g}\text{\/\hspace{0.03in}}%
\left( 1-gn_{k}\right) ^{g}}{n_{k}}=e^{x_{k}}
\end{equation}
where 
\begin{equation}
x_{k}=\beta \left( \varepsilon _{k}-\mu \right) .
\end{equation}
Eq.(3) is known as Haldane statistics[4].

Let us write Eq.(3) in a more compact form 
\begin{equation}
\frac{1}{y}\left( 1-y\right) ^{g}=e^{x}
\end{equation}
where y is defined as 
\begin{equation}
y=\frac{1}{\frac{1}{n_{k}}+1-g}.
\end{equation}
In order to get a real x on the right hand side of Eq.(5), $y\leq 1$. This
inequality imposes a condition 
\begin{equation}
gn_{k}\leq 1.
\end{equation}
Note that for g=1 inequality(7) gives $n_{k}\leq 1$ , for g=0; $n_{k\text{ }%
} $can take any values and for the intermediate values $n_{k\text{ }}$is
restricted as $n_{k}\leq \frac{1}{g}.$ Therefore, inequality(7) could be
interpreted as the generalization of \ the \textit{Pauli Exclusion principle.%
}

Since it is difficult to solve Eq.(5), an approximation is needed for y .
Expanding the factor $(1-y)^{g}$ in a Taylor series about y =0: 
\begin{equation}
\left( 1-y\right) ^{g}\approx 1-gy+O\left( y^{2}\right) .
\end{equation}
Thus, Eq.(5) becomes 
\begin{equation}
n_{k}=\frac{1}{e^{\beta \left( \varepsilon _{k}-\mu \right) }+2g-1}.
\end{equation}
The last approximation is a linear approximation to Eq.(3) and therefore
does not satisfy inequality (7), if g is not close to 0 or 1. Interesting
enough, Eq.(9) turns out to be the Maxwell distribution when $g=\frac{1}{2}$
[1].

\subsection{ Steepest Descent Method}

In order to get another solution of distribution function $n_{k}$ we are
willing to calculate it using the steepest descent optimization method [5].
Two of the authors of the present study FB and DD recently introduced a
partition function $\mathcal{Z}_{g}$ for a \ grandcanonical ensemble of a
g-on gas [1]. Let us write it for the extensive systems as 
\begin{equation}
\mathcal{Z}_{g}=\prod_{k=1}^{K}\left\{ \sum_{k=0}^{1}e^{-x_{k}}+\left(
1-g\right) \sum_{k=2}^{\infty }e^{-x_{k}}\right\} 
\end{equation}
where $x_{k}$ is given by Eq.(4). The given partition function leads to the
following partition function for the canonical ensemble

\begin{equation}
Z_{g}=\prod_{k=1}^{K}\left( 1+y_{k}\right) +\left( 1-g\right) \left(
y_{k}^{2}+y_{k}^{3}+...\right) \qquad \qquad
\end{equation}
where $y_{k}=e^{-\beta \varepsilon _{k}}$ and k denoting the quantum state $%
\varepsilon _{k}$. Here g is a parameter that unifies the partition
functions of boson and fermion systems under g-on gas system. For g=0 the
partition function in Eq.(11) becomes the partition function of a boson
system, for g=1 it becomes that of a fermion system. Let us find the
distribution function of a g-on system using this partition function and the
steepest descent method.

In order to use the steepest descent method one has to express the partition
function as a complex integral around a closed contour. For this purpose let
us define the function

\QTP{Body Math}
\begin{equation}
f_{g}\left( z\right) =\prod_{k=1}^{K}\left\{ \left( 1+zy_{k}\right) +\left(
1-g\right) \left( z^{2}y_{k}^{2}+z^{3}y_{k}^{3}+...\right) \right\} .
\end{equation}

Using this function, the partition function given in Eq.(11) can be written
as

\QTP{Body Math}
\begin{equation}
Z_{g}=\dfrac{1}{2\pi i}\oint\limits_{C}\dfrac{f_{g}\left( z\right) }{z^{N+1}}%
dz\qquad
\end{equation}

or $\qquad $

\begin{equation}
Z_{g}=\frac{1}{2\pi i}\oint\limits_{C}\prod_{k=1}^{K}\left[ \left( \frac{%
1+zy_{k}}{z^{N+1}}+\frac{\left( 1-g\right) z^{2}y_{k}^{2}\left(
1+zy_{k}+z^{2}y_{k}^{2}+...\right) }{z^{N+1}}\right) \right] \,dz.
\end{equation}
This integral (or partition function) can be approximated by using the
method of steepest descent. In order to see it let us write

\begin{equation}
e^{h\left( z\right) }=\frac{f_{g}\left( z\right) }{z^{N+1}}
\end{equation}
where $f_{g}\left( z\right) $ is given by Eq.(12). Using \ this definition
one can rewrite the $h(z)$ as

\begin{equation}
h\left( z\right) =\ln \prod_{k=1}^{K}\left[ \left( \frac{1+zx_{k}}{z^{N+1}}+%
\frac{\left( 1-g\right) z^{2}y_{k}^{2}\left(
1+zy_{k}+z^{2}y_{k}^{2}+...\right) }{z^{N+1}}\right) \right]
\end{equation}
or as 
\begin{equation}
h(z)=\dsum_{k=1}^{\infty }\left\{ \ln \left[ \left( 1+zy_{k}\right) +\left(
1-g\right) z^{2}y_{k}^{2}\left( 1+zy_{k}+z^{2}y_{k}^{2}+...\right) \right]
\right\} -\ln z^{-N+1}.
\end{equation}

The function $h(z)$ has a minimum at $z_{o}$ and this minimum is very steep
because the function $f_{g}\left( z\right) $ consists of infinite number of
multiplication and $N+1$ is very large. Thus by using the approximation
formula of the steepest descent method, one gets an expression for the
partition function of a g-on gas as[6]

\begin{equation}
Z_{g}\cong \left[ z_{o}^{\_N+1}f_{g}\left( z_{0}\right) 2\pi f^{\prime
\prime }\left( z_{0}\right) \right] ^{1/2}.
\end{equation}
In order to find the distribution function one applies the steepest descent
condition to h(z), that is 
\begin{equation}
h^{\prime }(z_{0})=-\frac{N+1}{z_{0}}+\sum_{k=0}^{\infty }\frac{y_{k}+\left(
1-g\right) \left( 2z_{0}y_{k}^{2}+3z_{0}^{2}y_{k}^{3}+...\right) }{%
1+z_{0}y_{k}+\left( 1-g\right) z_{0}^{2}y_{k}^{2}\left(
1+z_{0}y_{k}+z_{0}^{2}y_{k}^{2}+...\right) }=0.
\end{equation}
Since $\ N\gg 1,$ one can write $N+1\approx N.$ Using this approximation and
arranging Eq.(18) one obtains 
\begin{equation}
N\approx \sum_{k=0}^{\infty }\frac{z_{0}y_{k}+\left( 1-g\right) \left(
2z_{0}^{2}y_{k}^{2}+3z_{0}^{3}y_{k}^{3}+...\right) }{1+z_{0}y_{k}+\left(
1-g\right) z_{0}^{2}y_{k}^{2}\left(
1+z_{0}y_{k}+z_{0}^{2}y_{k}^{2}+...\right) }\ .
\end{equation}
The factor $\frac{1}{z_{0}}$ can be interpreted as fugacity $e^{-\beta \mu }$
[6]. For the sake of simplicity let us define a new variable as 
\begin{equation}
a_{k}=e^{-\beta \left( \varepsilon _{k}-\mu \right) }=e^{-x_{k}}=z_{0}y_{k}.
\end{equation}
Using the series 
\begin{equation}
1+a_{k}+a_{k}^{2}+a_{k}^{3}+...=(1-a_{k})^{-1}
\end{equation}
\begin{equation}
1+2a_{k}+3a_{k}^{2}+...=(1-a_{k})^{-2}
\end{equation}
where $a_{k}<1$. By means of these series and the definition (21) one can
write Eq.(20) as 
\begin{equation}
N=\sum_{k=1}^{\infty }\frac{a_{k}}{a_{k}+\frac{ga_{k}^{2}-2a_{k}+1}{%
1-2ga_{k}+ga_{k}^{2}}}=\sum_{k=1}^{\infty }\frac{a_{k}\left(
1-2ga_{k}+ga_{k}^{2}\right) }{\left( 1-a_{k}\right) \left(
1-ga_{k}^{2}\right) }.
\end{equation}
or 
\begin{equation}
N=\sum_{k=1}^{\infty }\frac{e^{-x_{k}}\left(
1-2ge^{-x_{k}}+ge^{-2x_{k}}\right) }{\left( 1-e^{-x_{k}}\right) \left(
1-ge^{-2x_{k}}\right) }.
\end{equation}
Since the total number of particles of a system is given as 
\begin{equation}
N=\sum_{k=1}^{\infty }\overline{n_{k}},
\end{equation}
the terms in Eq.(25) gives the number of particles which occupy the k$^{th}$
state. Hence for the distribution of g-on particles it is possible to write 
\begin{equation}
\overline{n_{k}}=\frac{e^{-x_{k}}\left( 1-2ge^{-x_{k}}+ge^{-2x_{k}}\right) }{%
\left( 1-e^{-x_{k}}\right) \left( 1-ge^{-2x_{k}}\right) }.
\end{equation}
Eq.(27) which unifies the distribution of the quantum gases recovers the
well known Bose and Fermi distributions for g=0 and g=1 respectively.

In order to find a more elegant result which could be used moreeasily in the
calculations the term in Eq.(24) can be written as 
\begin{equation}
\frac{ga_{k}^{2}-2a_{k}+1}{1-2ga_{k}+ga_{k}^{2}}\approx 1+(2g-2)a_{k.}
\end{equation}
By taking into account Eq.(28), Eq.(27) reads, 
\begin{equation}
\overline{n_{k}}=\frac{a_{k}}{1+a_{k}\left( 2g-1\right) }
\end{equation}
which is valid about g=1 and g=0. Let us write Eq.(29) explicitly;$\qquad
\qquad \qquad \qquad \qquad \qquad \qquad $ 
\begin{equation}
\overline{n_{k}}=\frac{1}{e^{\beta \left( \varepsilon _{k}-\mu \right) }+2g-1%
}.
\end{equation}

It is observed that Eq.(30) is same as Eq.(9) which is obtained through an
approximation to Haldane formula.

\subsection{ Statistical Method}

The partition function of g-ons which is given by Eq.(10) can be used to
find the probability of the k$^{th}$ state being occupied by $n_{k}$ g-ons.
If the state k with an energy $\varepsilon _{k}$ is occupied by $n_{k}$
g-ons then the other states are occupied by N-$n_{k}$ g-ons, where N is the
total particle number. Therefore the ratio of the partition function for the
states except k$^{th}$ state times $e^{-n_{k}x_{k}}$ to the partition
function gives us the probability of finding $n_{k\text{ }}$particles in the
k. state. In the thermodynamic limit i.e. as $N\rightarrow \infty $ this
probability can be written as 
\begin{equation}
P_{g}^{k}\left( n_{k}\right) =\frac{e^{-n_{k}x_{k}}\dprod\limits_{\substack{ %
i=1  \\ i\neq k}}\left\{ \sum\limits_{n_{i}=0}^{1}e^{-n_{i}x_{i}}+\left(
1-g\right) \sum\limits_{n_{i}=2}^{\infty }e^{-n_{i}x_{i}}\right\} }{%
\dprod\limits_{i=1}\left\{ \sum\limits_{n_{i}=0}^{1}e^{-n_{i}x_{i}}+\left(
1-g\right) \sum\limits_{n_{k}=2}^{\infty }e^{-n_{i}x_{i}}\right\} }.
\end{equation}
or writing in a simplified form: 
\begin{equation}
P_{g}^{k}\left( n_{k}\right) =\frac{e^{-n_{k}x_{k}}}{\sum%
\limits_{n_{k}=0}^{1}e^{-n_{k}x_{i}}+\left( 1-g\right)
\sum\limits_{n_{k}=2}^{\infty }e^{-n_{k}x_{i}}}.
\end{equation}
After performing the summation Eq.(32) turns out to be 
\begin{equation}
P_{g}^{k}\left( n_{k}\right) =\frac{e^{-n_{k}x_{k}}\left(
1-ge^{-x_{k}}\right) }{1-ge^{-2x_{k}}}.
\end{equation}
It could be mentioned that Eq.(33) is a unified form of the probabilities of
occupation number of the quantum gases simply bosons and fermions. One may
write the distribution functions by means of Eq.(33) for g-ons as 
\begin{equation}
\overline{n_{k}}=\sum_{n_{k}=0}^{1}n_{k}P_{g}^{k}\left( n_{k}\right) +\left(
1-g\right) \sum_{n_{k}=2}^{\infty }n_{k}P_{g}^{k}\left( n_{k}\right) .
\end{equation}
After performing the summations it reads 
\begin{equation}
\overline{n_{k}}=\frac{e^{-x_{k}}\left( 1-e^{-x_{k}}\right) }{1-ge^{-2x_{k}}}%
\left[ g+\frac{1-g}{\left( 1-e^{-x_{k}}\right) ^{2}}\right] =\frac{%
e^{-x_{k}}(ge^{-2x_{k}}-2ge^{-x_{k}}+1)}{\left( 1-ge^{-2x_{k}}\right) \left(
1-e^{-x_{k}}\right) }
\end{equation}
or more explicitly; 
\begin{equation}
\overline{n_{k}}=\frac{e^{-\beta \left( \varepsilon _{k}-\mu \right) }\left(
1-2ge^{-\beta \left( \varepsilon _{k}-\mu \right) }+ge^{-2\beta \left(
\varepsilon _{k}-\mu \right) }\right) }{\left( 1-e^{-\beta \left(
\varepsilon _{k}-\mu \right) }\right) \left( 1-ge^{-2\beta \left(
\varepsilon _{k}-\mu \right) }\right) }
\end{equation}
which is the same as Eq.(27) as expected.

\section{Fluctuations in the occupation number}

In the preceding section we have calculated the probability $P_{g}^{k}\left(
n_{k}\right) $ of occupying a g-on of the k$^{th}$ state in the Fock space
and the distribution function $\overline{n_{k}}$ .

The deviations of mean values in the distribution function are given by the
standard deviations of distributions, 
\begin{equation}
\sigma ^{2}=\overline{n_{k}^{2}}-\overline{n_{k}}^{2}.
\end{equation}
Therefore at first $\overline{n_{k}^{2}}$ has to be calculated. For this
purpose let us write the following equation 
\begin{equation}
\overline{n_{k}^{2}}=\sum_{n_{k}=0}^{1}n_{k}^{2}P_{g}^{k}\left( n_{k}\right)
+\left( 1-g\right) \sum_{n_{k}=2}^{\infty }n_{k}^{2}P_{g}^{k}\left(
n_{k}\right) .
\end{equation}
Substituting the expression for $P_{g}^{k}\left( n_{k}\right) $ into Eq.(38)
and performing the summations one finds 
\begin{equation}
\overline{n_{k}^{2}}=\frac{e^{-x_{k}}\left( 1-e^{-x_{k}}\right) }{%
1-ge^{-2x_{k}}}\left[ g+\left( 1-g\right) \frac{1+e^{-x_{k}}}{\left(
1-e^{-x_{k}}\right) ^{3}}\right] .
\end{equation}
By means of Eqs.(35),(37) and (38) and one can calculates the variance as 
\begin{equation}
\sigma ^{2}=\frac{%
e^{-x_{k}}(g^{2}e^{-4x_{k}}-4ge^{-3x_{k}}+6ge^{-2x_{k}}-4ge^{-x_{k}}+1)}{%
\left( 1-e^{-x_{k}}\right) ^{2}\left( 1-ge^{-2x_{k}}\right) ^{2}}.
\end{equation}
and the ratio $\frac{\sigma ^{2}}{\overline{n_{k}}^{2}}$ as 
\begin{equation}
\frac{\sigma ^{2}}{\overline{n_{k}}^{2}}=\frac{%
e^{x_{k}}(g^{2}e^{-4x_{k}}-4ge^{-3x_{k}}+6ge^{-2x_{k}}-4ge^{-x_{k}}+1)}{%
\left( ge^{-2x_{k}}-2ge^{-x_{k}}+1\right) ^{2}}
\end{equation}
which reduces to the standard results.

\section{Thermodynamics of the g-on gas}

In this section, as a concrete application of g-on statistics we would like
to investigate the thermodynamics of an indistinguishable g-on gas. The
ideal g-on gas is a model system, where the influence of quantum statistical
effects can be very well studied. Our aim is to calculate the
thermodynamical potential of the g-on gas $\Phi _{g}(T,V,\mu )$. 
\begin{equation}
\Phi _{g}\left( T,V,\mu \right) =-k_{B}T\ln \mathcal{Z}_{g}\left(
T,V,z\right) 
\end{equation}
where summation approximately is over all of the quantum states. By taking
into account equation (10) one derives approximately 
\begin{equation}
\Phi _{g}\left( T,V,\mu \right) =-k_{B}T\sum_{k=1}^{K}\ln \left[ 1+\frac{1}{%
\omega _{k}}\right] 
\end{equation}
where 
\begin{equation}
\omega _{k}\cong \frac{1}{e^{-x_{k}}+g-1}.
\end{equation}
Since it is difficult to handle with exact formula for $\overline{n_{k}}$
its approximation Eq.(10) or Eq.(30) is appropriate 
\begin{equation}
\overline{n_{k}}=\frac{1}{\omega _{k}+g}.
\end{equation}
Given the generalized thermodynamic potential 
\begin{equation}
\Phi _{g}\left( T,V,\mu \right) =-p_{g}V
\end{equation}
the pressure of the g-on gas is obtained as 
\begin{equation}
p_{g}V=k_{B}T\sum_{k=1}^{K}\ln \left[ 1+\frac{1}{\omega _{k}}\right] 
\end{equation}
while the total number of particles is found as 
\begin{equation}
N_{g}=\sum_{k}\frac{1}{\omega _{k}+g}.
\end{equation}
The state of the gas is determined by Eqs. (47) and (48). But one can also
calculate the energy of the gas which is given by 
\begin{equation}
E_{g}=\sum_{k}\overline{n_{k}}\,\varepsilon _{k}=\sum_{k}\frac{\varepsilon
_{k}}{\omega _{k}+g}.
\end{equation}
The logarithm of the grandpartition function $Q_{g}$($T$,$V$,$z$) is given
as 
\begin{equation}
Q_{g}(T,V,z)=\ln \mathcal{Z}_{g}\left( T,V,z\right) =\sum_{k=1}^{K}\ln \left[
1+\frac{1}{\omega _{k}}\right] .
\end{equation}
Let one of the g-on energies $\varepsilon _{k\text{ }}$are those of the
energies of a free quantum mechanical particle in a box of volume V. The
chemical potential or fugacity is not fixed, but the total number of g-ons
are conserved on the average and z has to be determined from Eq.(48). For a
large volume the sum over all single g-on states can be given as the single
g-on density states, where spin is not taken into account. Hence, 
\begin{equation}
D\left( \varepsilon \right) =A\varepsilon ^{1/2}
\end{equation}
can be written where 
\begin{equation}
A=\frac{2\pi V}{h^{3}}\left( 2mh\right) ^{3/2}.
\end{equation}
Thus the summation in Eq.(50) becomes 
\begin{equation}
Q_{g}(T,V,z)=\int\limits_{0}^{\infty }\ln \left[ 1+\frac{1}{\omega _{k}}%
\right] \varepsilon ^{1/2}d\varepsilon 
\end{equation}
where Eq.(51) is taken into account. When the integration is performed by
parts, Q(T,V,z) is reduced to: 
\begin{equation}
Q_{g}(T,V,z)=\frac{2}{3}A\beta \int\limits_{0}^{\infty }\varepsilon
^{3/2}nd\varepsilon .
\end{equation}
After substituting n from Eq.(45) and then rearranging 
\begin{equation}
Q_{g}(T,V,z)=\frac{2}{3}A\beta z\int\limits_{0}^{\infty }\frac{\varepsilon
^{3/2}}{e^{\beta \varepsilon }-\left( 1-2g\right) z}d\varepsilon 
\end{equation}
is obtained, where z is the usual fugacity i.e. $z=e^{\beta \mu }.$ The
integral is calculated using the related formula pp. 326 of reference[7].
Thus one concludes that for a g-on system 
\begin{equation}
Q_{g}(T,V,z)=\frac{2}{3}A\beta zF_{5/2}\left( g,z\right) 
\end{equation}
where the definition 
\begin{equation}
F_{n}\left( g,z\right) =\frac{\Gamma \left( n\right) }{z\left( 1-2g\right)
\beta ^{n}}\sum_{k=1}^{K}\frac{\left[ z\left( 1-2g\right) \right] ^{k}}{k^{n}%
}
\end{equation}
is adopted in which $\Gamma \left( n\right) $ is the usual gamma function.
Then the state of the g-on system is determined by the following equation: 
\begin{equation}
p_{g}V=k_{B}TQ_{g}\left( T,V,z\right) .
\end{equation}
Thus, the equation of state of fermions and bosons is unified

In thermodynamics limit, that is in the limit of infinite volume with
particle density held fixed,particles in the ground state, having no
contribution of the kinetic energy to the pressure, then, in terms of the
thermal wavelength, the pressure of the Bose system becomes, as it is
expected, 
\begin{equation}
p_{0}=\frac{kT}{\lambda ^{3}}\sum_{k=1}^{\infty }\frac{z^{k}}{k^{5/2}}
\end{equation}
where $\Gamma \left( \frac{5}{2}\right) =3\frac{\sqrt{\pi }}{4}$ is taken
and $\lambda $is the thermal wavelength [8].

In a similar manner,the thermodynamical properties of a Fermi system follows
immediately from the logarithm of the grandpartition function of a g-on
system, which is given by Eq. (43). In terms of the thermal wavelength, the
pressure of the Fermi system is 
\begin{equation}
p_{1}=g_{s}\frac{k_{B}T}{\lambda ^{3}}\sum_{k=1}^{\infty }\frac{\left(
-1\right) ^{k-1}z^{k}}{k^{5/2}}
\end{equation}
where $g_{s}$ is a factor due to spin [8].

For a large volume, in accordance with equation (49), the total number of
g-ons could be written as 
\begin{equation}
N_{g}\left( T,V,z\right) =Az\int\limits_{0}^{\infty }\frac{\varepsilon ^{1/2}%
}{e^{\beta \varepsilon }-\left( 1-2g\right) z}d\varepsilon 
\end{equation}
assuming that single number of g-on states can be calculated in terms of
integrals. Calculation of the integral in Eq.(61) using the related
integrals in reference[7] pp326 leads to 
\begin{equation}
N_{g}\left( T,V,z\right) =AzF_{3/2}\left( g,z\right) 
\end{equation}
where $F_{3/2}\left( g,z\right) $ is defined by Eq.(58) [8]. If \ g=0 and
g=1 are substituted in Eq.(62) the formula for the total number of Bose
system $N_{0}\left( T,V,z\right) $ and Fermi system $N_{1}\left(
T,V,z\right) $ are respectively recovered .

For a Bose system the total number of particles $N_{0}\left( T,V,z\right) $
including $\varepsilon =0$ energy states, is found to be: 
\begin{equation}
N_{0}\left( T,V,z\right) =AzF_{3/2}(z)+\frac{z}{1-z}
\end{equation}
where the last term represents the contribution of the energy level $%
\varepsilon =0$ to the mean particle number.

In a similar manner, for the total number of fermions one gets 
\begin{equation}
N_{1}\left( T,V,z\right) =g_{s}\frac{V}{\lambda ^{3}}\sum_{k=1}^{\infty }%
\frac{\left( -1\right) ^{k-1}z^{k}}{k^{1/2}}
\end{equation}
where $\Gamma \left( 3/2\right) =\sqrt{\pi /2}$ is taken. This is an
expected standard result [8]\-

\bigskip

\bigskip {\huge Conclusions}

This study has been initiated from the point of view bosons and fermions,
which obey quantum statistics, could be regarded as g-ons which obey
fractional statistics. In this paper, Haldane statistics is rederived by
variational method where entropy of the g-on gas is involved. Due to the
importance of the distribution function, by means of steepest descent and
statistical methods, mean occupation number is derived where partition
function of g-on gas is used. It is concluded that mean occupation number
which is found by these methods lead to same expressions. Within this
context the probabilities of occupation number of g-on gas is also
calculated which unifies the probabilities of quantum gases.

The approximate solutions of the distribution function of a g-on gas system,
which is found from the variational, steepest descent and statistical
methods are developed to an elegant expression that recovers FD distribution
for g=1, and BE distribution for g=0. It should be remarked that the
partition function which is used in this study is appropriate for the g-on
gas since it leads to the same approximate expression for the mean
occupation number with Haldane statistics.

As an application of the mean occupation number, the thermodynamical
formulation of the g-on gas is derived which gives the opportunity to unify
the thermostatistical formulation of quantum gases, simply fermions and
bosons.

\bigskip

Authors would like to thank Ege University Research Fund for their partial
support under the Project Number 2000 Fen 19.

\bigskip

{\huge References}

\begin{enumerate}
\item  B\"{u}y\"{u}kk\i l\i \c{c} F., Demirhan D., Eur.Phys.Jour.B. 14
(2000) 704.

\item  G.Su, M.Suzuki, Eur Phys.Jour.B. 5\ (1998) 577.

\item  Y.Su Wu, Phys.Rev.Lett.73 (1994) 922

\item  F.D.M. Haldane, Phys.Rev.Lett. 66 (1991) 1529.

\item  C.B.Arfken, H.J.Weber, \textit{Mathematical Methods for Physicists}
(Academic Press) 1966

\item  C.Kittel, \textit{Elementary Statistical Physics}$\mathbb{\ }$(John
Wiley, New York) 1958.

\item  I.S.Grandshteyn, I.M.Ryzhik, \textit{Table of Integral Series and
Products }(Academic Press, New York) 1965.

\item  W.Greiner, L. Neise, H Stoecker, \textit{Thermodynamics and
Statistical Mechanics }(Springer Verlag) 1995.
\end{enumerate}

\end{document}